\documentclass{elsart}

\usepackage{natbib}

 \usepackage{graphicx}

\usepackage{amssymb}

\journal{New Astronomy}

\begin{document}

\begin{frontmatter}



\title{Drift Wave Model of Rotating Radio Transients}


\author[Purdue1]{D.Lomiashvili\corauthref{cor1}}
\ead{lomiashvili@gmail.com}
 \corauth[cor1]{Corresponding author.}
\author[Tbilisi1]{G.Machabeli}
\ead{g\_machabeli@hotmail.com}
\author[Pushchino1]{I.Malov}
\ead{malov@prao.ru}

\address[Purdue1]{Purdue University, 525 Northwestern
              Avenue, West Lafayette, IN, USA}
\address[Tbilisi1]{Tbilisi state university, 3 Chavchavadze Ave., 0128,
              Tbilisi, Georgia}
\address[Pushchino1]{Pushchino Radio Astronomy Observatory, P.N.Lebedev
              Institute of Physics, 142290, Pushchino, Moscow Region, Russia}

\begin{abstract}
During the last few years there were discovered and deeply
examined several transient neutron stars ("Rotating Radio
Transients"). It is already well accepted that these objects are
rotating neutron stars. But their extraordinary features
(burst-like behavior) made necessary revision of well accepted
models of pulsar interior structure. Nowadays most popular model
for RRATs is precessing pulsar model, which is the subject of big
discussion. We assume that these objects are pulsars with specific
spin parameters. An important feature of our model, naturally
explaining most of the properties of these neutron stars, is
presence of very low frequency, nearly transverse drift waves
propagating across the magnetic field and encircling the open
field lines region of the pulsar magnetosphere.
\end{abstract}


\end{frontmatter}

\section{Introduction}
\label{intro} Recently, \cite{Mcl05} reported the discovery of a
new class of radio transients from the Parkes Multibeam Pulsar
Survey. The current sample includes 11 objects characterized by
single, dispersed bursts of radio emission with the durations
ranging from 2 to 30 milliseconds. Long-term monitoring of these
objects led to identification of their spin periods (P), ranging
from 0.4 to 7 seconds. As \citet{Mcl05} concluded, these objects
represent a previously unknown population of rotation-powered
neutron stars, which were named as Rotating RAdio Transients
(RRATs). Another interesting case between normal pulsars and RRATs
have been reported earlier (PSRs J1649+2533 and J1752+2359) by
\citet{Lew04}. Understanding the physical origin of these objects
as well as their relationship with normal pulsars is desirable.
There exist few models for this phenomenon: 1) The precession
model; 2) The model suggesting that these objects are pulsars
located slightly below the radio emission "death line", and become
active occasionally when the conditions for pair production and
coherent emission are satisfied; 3) The third model invoking a
radio emission direction reversal in normal pulsars. In this
picture, our line of sight misses the main radio emission beam of
RRATs but happens to sweep the emission beam when the radio
emission direction is reversed. Last two ideas were suggested by
\cite{Zhang06}. Here we propose another model of RRATs proving
once again that they are normal pulsars with special values of
certain parameters.

The paper is organized as follows. Pulsar radio emission mechanism
is presented in Section 2. The generation of the drift waves and
their influence on the curvature of magnetic field lines are
discussed in Section 3. Our proposed model is presented in Section
4. The conclusions are summarized in Section 5.

\section{Emission mechanism}
As it is known the pulsar magnetosphere is filled by a dense
relativistic electron-positron plasma. The (e$^{+}$e$^{-}$) pairs
are generated as a consequence of the avalanche process (first
described by \citet{sturrock71}) and flow along the open magnetic
field lines. The plasma is multi-component, with a one-dimensional
distribution function ( see Fig.1 from  \citet{arons81}) and
consists of the following components: the bulk of plasma with an
average Lorentz-factor $\gamma_{p}\simeq10$; a tail on the
distribution function with $\gamma_{t}\simeq10^{4}$ and the
primary beam with $\gamma_{b}\simeq10^{6}$. The main mechanism of
wave generation in plasmas of the pulsar magnetosphere is the
cyclotron instability. Generation of waves is possible if the
condition of the cyclotron resonance if fulfilled \citep{kaz91a}:
\begin{equation}
    \omega-k_{\varphi}V_{\varphi}-k_{x}u_{x}+\frac{\omega_{B}}{\gamma_{r}}=0,
\end{equation}
where $V_{\varphi}$ is the particle velocity along the magnetic
field, $\gamma_{r}$ is the Lorentz-factor for the resonant
particles and $u_{x}=cV_{\varphi}\gamma_{r}/\rho\omega_{B}$ is the
drift velocity of the particles due to curvature of the field
lines ($\rho$ is the radius of curvature of the field lines and
$\omega_{B}=eB/mc$ is the cyclotron frequency). Here cylindrical
coordinate system is chosen, with the $x$-axis directed
transversely to the plane of field line, when $r$ and $\varphi$
are the radial and azimuthal coordinates. Generated waves leave
the magnetosphere propagating at very small angles to the pulsar
local magnetic field lines and reach an observer as pulsar radio
emission. These processes take place near the light cylinder where
the cyclotron instability occurs \citep{lyut99}.

\section{Change of field line curvature and emission direction by drift waves}
It has been shown by \cite{kaz91b,kaz96} that, in addition to the
radio waves, very low frequency, nearly transverse drift waves can
be excited in the same region. The period of the drift waves
$P_{dr}$ can be written as:
\begin{equation}
P_{dr}=\frac{e}{4\pi ^{2}mc}\frac{BP^{2}}{\gamma}
\end{equation}
Where $P$ is the pulsar spin period, $B=B_{s}(R_{0}/R)^{3}$ is the
magnetic field in the wave excitement region and $\gamma $ is the
relativistic Lorentz factor of the particles. It appears that the
period of the drift wave can vary in a broad range. The magnetic
field of drift wave adds with pulsar magnetic field as $r$
component and causes changing of field line curvature $\rho _{c}$.
Here and below the cylindrical coordinate system ($x,r,\varphi $)
is chosen, with the $x$-axis directed transversely to the plane of
field line, while $r$ and $\varphi$ are the radial and azimuthal
coordinates, respectively.

Even a small change of $B_{r}$ causes significant change of $\rho
_{c}$. Variation of the field line curvature can be estimated as
\begin{equation}
\frac{\Delta \rho }{\rho }\approx k_{\varphi }r\frac{\Delta B_{r}}{
B_{\varphi }}
\end{equation}
Here $k_{\varphi }$ is a longitudinal component of wave vector and
$r$ is distance to the center of pulsar. It follows that even the
drift wave with a modest amplitude $B_{r}\sim \Delta B_{r}\sim
0.01B_{\varphi }$ alters the field line curvature substantially,
$\Delta \rho /\rho \sim 0.1$

Since radio waves propagates along the local magnetic field lines,
curvature variation causes change of emission direction.

\section{The model}

There is unequivocal correspondence between the observable intensity and $%
\alpha $ (angle between observers line and emission direction (see
fig. 1)). Maximum of intensity corresponds to minimum of $\alpha
$. The period of pulsar is time interval between neighboring
maxima of observable intensity i.e. minima of $\alpha $ (see fig.
2). According to this fact, we can say that the observable period
depends on time behavior of $\alpha $ and as it will appear below
it might differ from the spin period of pulsar.

\begin{figure}
\includegraphics[width=1 \textwidth]{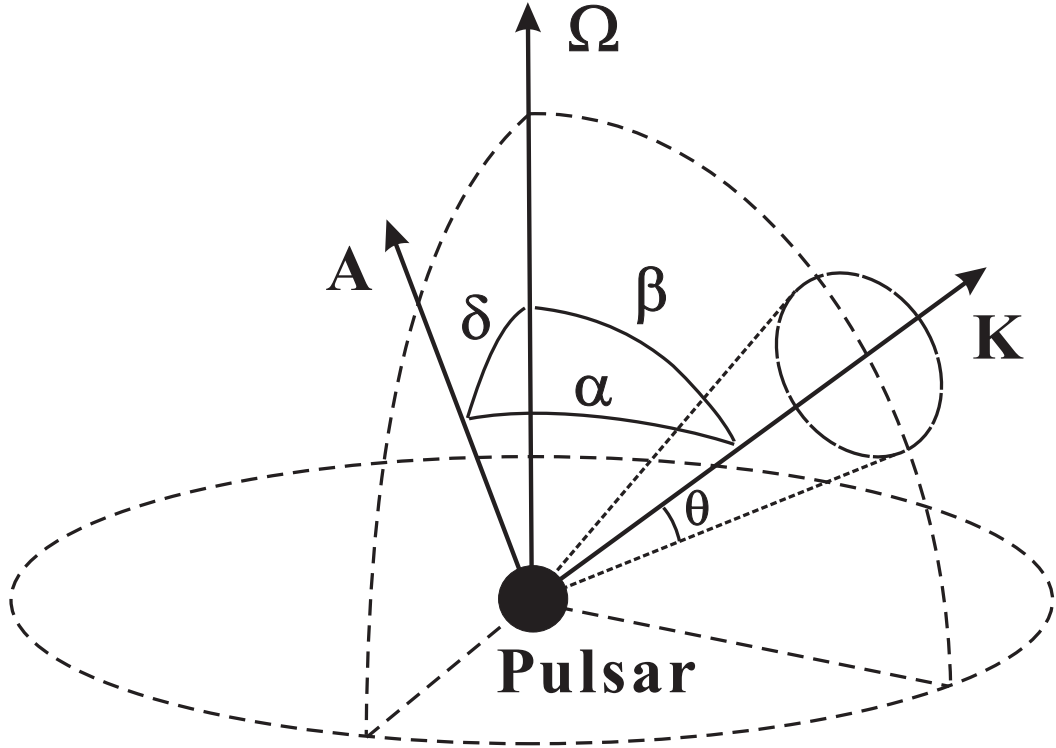} \caption{Geometry of {\bf $\Omega$} -
rotation, {\bf K} - emission and {\bf A} - observers axes. Angles
$\delta$ and $\vartheta $ are constants, while $\beta$ and
$\alpha$ are oscillating with time.}
\end{figure}

\begin{equation}
\cos \alpha = {\bf A}\cdot{\bf K}
\end{equation}
${\bf A }$ and ${\bf K}$ are unit guide vectors of observers and
emission axes respectively. In the spherical coordinate system
($r,\varphi ,\theta $), combined with plane of pulsar rotation,
these vectors can be expressed as:
\begin{equation}
{\bf A}=\left( 1,0,\delta \right)
\end{equation}
\begin{equation}
{\bf K}=\left( 1,\Omega t,\beta \right)
\end{equation}
where $\Omega =2\pi/P$ is the angular velocity of the pulsar.
$\delta $ is the angle between rotation and observers axis, and
$\beta $ is the angle between rotation and emission axis (see fig.
1). From equations (4),(5) and (6) follows that:
\begin{equation}
\alpha =\arccos \left( \sin \delta \sin \beta \cos \Omega t+\cos
\delta \cos \beta \right)
\end{equation}

In the absence of the drift wave $\beta =\beta _{0}=const$ and
consequently the period of $\alpha $ equals to $ 2\pi/\Omega $,
while in the case of existence of drift wave $\beta$ is
oscillating with time \citep{lom06}.

Considering that for wavelengths of order of the transverse scale
of magnetosphere, the treatment of the drift waves as plane waves
becomes inappropriate the geometry of magnetosphere must be taken
into account. A treatment in terms of spherical harmonic
eigenmodes with specific $l,m$ is then appropriate \citep{gog05}.
Therefore in general case the amplitude $\Delta \beta$ must be
defined as follows

\begin{equation} \Delta \beta =\sum \Delta
\beta_{n} \sin \left( \omega^{n} _{dr}t+\varphi^{n} \right)
\end{equation}

Here $n$ is number of eigenmode. According to equations (7) and
(8) we obtain

\begin{eqnarray}
\alpha =\arccos [ \sin \delta \sin \left( \beta _{0}+\sum \Delta
\beta_{n} \sin \left( \omega^{n} _{dr}t+\varphi^{n} \right)
\right) \cos \Omega t\nonumber
 \\
+ \cos \delta \cos \left( \beta _{0}+\sum \Delta \beta_{n} \sin
\left( \omega^{n} _{dr}t+\varphi^{n} \right) \right) ]
\end{eqnarray}

\begin{equation}
\alpha _{\min }^{k}=\left| \left( \beta _{0}-\delta \right) +\Delta
\beta \sin \left( 2\pi k\frac{\omega _{dr}}{\Omega }+\varphi \right)
\right|
\end{equation}

Where $\alpha_{\min }^{k}$ is the minimum of $\alpha$ after $k$
revolutions of the pulsar, after time reckoning\footnote{as zero
point of time reckoning is taken detection moment of any pulse.}.
The parameters of the pulse profile (e.g. width, height etc.)
significantly depend on what the minimal angle would be between
emission axis and observers axis when the first one passes the
other (during one revolution). If the emission cone does not cross
the observers line of sight entirely (i.e. minimal angle between
them is more then cone angle $\vartheta$, see inequality (11))
then pulsar emission is unobservable for us. On the other hand,
inequality (12) defines condition that is necessary for emission
detection.

\begin{equation}
\alpha _{\min }^{k}>\vartheta
\end{equation}
\begin{equation}
\alpha _{\min }^{k}<\vartheta
\end{equation}

Hence for some values of parameters $\Omega$, $\omega_{dr}$, $\beta $, $%
\Delta \beta $, $\delta $, $\varphi $ and $\vartheta $ (Set A) it
is possible to accomplish following regime: Once condition (12)
becomes fulfilled it stays along $k$ revolutions, therefore pulsar
is observable during $k$ periods. While over the the other $m-k$
revolutions condition (11) accomplishes, consequently pulsar
appears "switched off" (see Fig. 2). It means that burst-like
emission occurs.

\begin{figure}
\includegraphics[width=1 \textwidth]{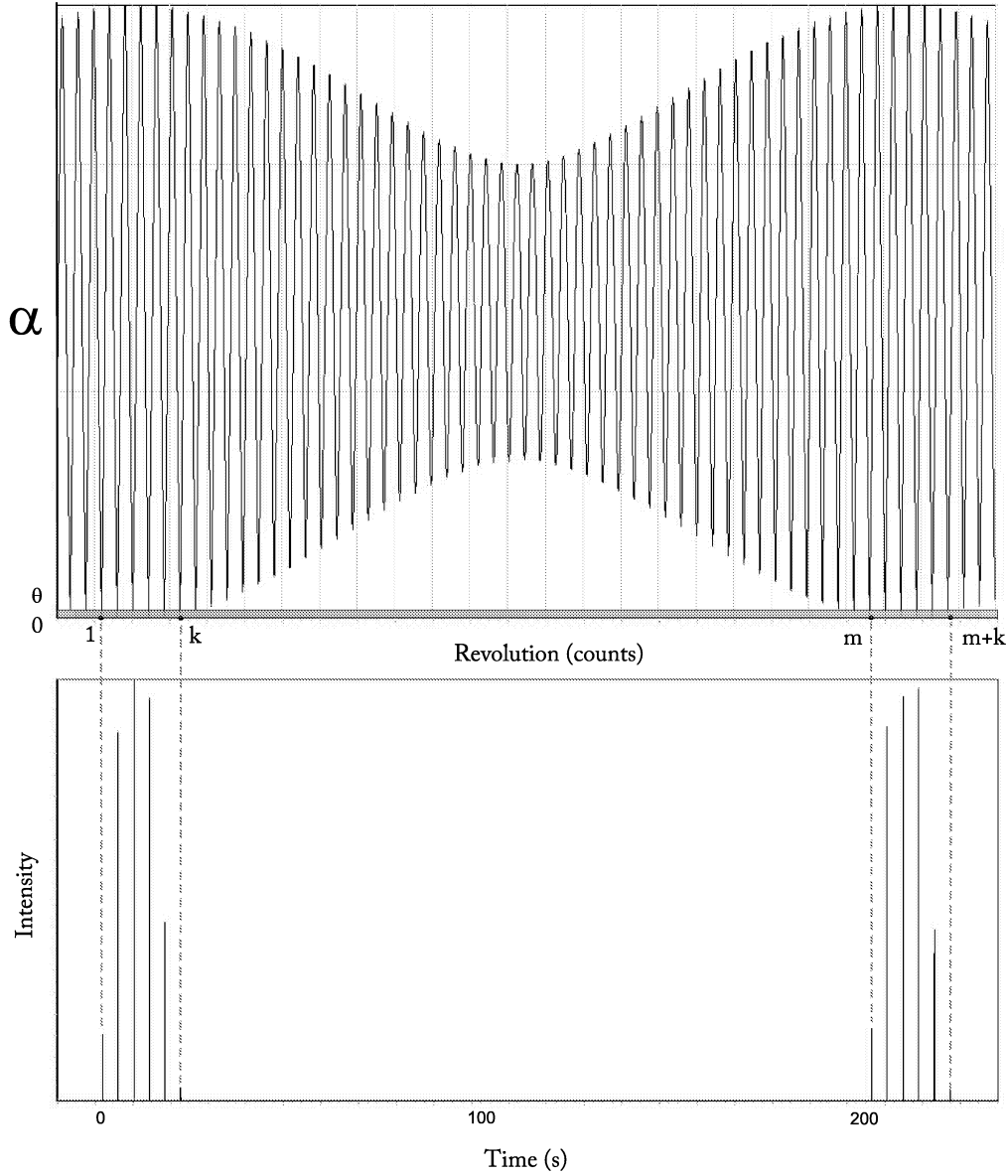} \caption{The oscillating behavior of $\alpha $ with time
\ for $\beta _{0}=1.35$, $\delta=1.0$, $\Delta \beta=0.35$,
$\omega _{dr}=2\pi(1/4.26+1/204.4)(s^{-1})$,
$\Omega=2\pi/4.26(s^{-1})$, $\varphi=1.5$ and Simulated lightcurve
of PSR J1819-1458}(below)
\end{figure}

PSR J1752+2359 has been selected for its unusually long nulling
periods. It had been observed on several occasions between 2000
and 2002 by \cite{Lew04}. The pulsar spends 70- 80\% of the time
in a "quasi-null" state. The "on-states" occur once every 400-600
periods and last, on average, for $\sim 100$ periods. A more
detailed inspection of the on-states reveals that these burst-like
emissions are quite similar in shape and duration and they decay
into a null state in a manner that is quite reasonably described
by a $te ^{-t/\tau}$ function with $\tau \sim 45$ seconds. PSR
J1752+2359 switches off gradually, rather than suddenly. Simulated
lightcurve of J1752+2359 is presented in Fig. 3.

\begin{figure}
\includegraphics[width=1 \textwidth]{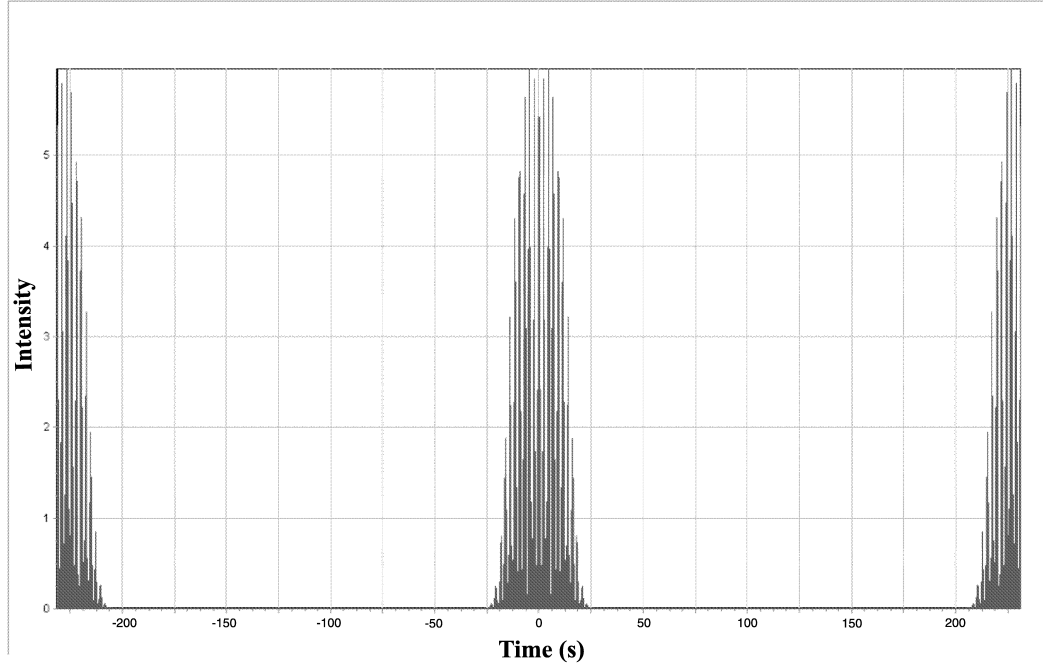} \caption{Simulated lightcurve
of PSR J1752+2359}
\end{figure}

If we consider these pulsars in the framework of our model their
parameters (spin, angular etc.) will get values shown in Table 1.
Simulated lightcurve of PSR J1819-1458 (RRAT J1819-1458) is
presented in Fig. 2.

\section{Conclusions}

To sum up, drift wave driven model is very convenient since it
allows to explain almost every extraordinary feature of known
pulsars such as: extremely long periods \citep{lom06}, cyclic
variations of radiation intensity (RRATs) and rotational
parameters (in prep.), subpulse drift \citep{gog05}, nullings
\citep{kaz96}, etc. Moreover it has potentiality to unfold future
discoveries in this field.

In long period radio pulsars \citep{lom06} relation between
periods of pulsar and drift wave along with pulsar geometry
satisfy much precise condition than in RRATs. Therefore total
number of RRATs should be much more than long period radio
pulsars.

The precession model is very similar to ours although it has
number of disadvantages. First of all, it is necessary to take
large value of wobble angle to interpret observational data of
some RRATs. In the case of neutron stars this is hard to explain
because of their superfluid interior structure. However, there is
a need of distinguishable sign to elucidate which model is true.
\cite{Zhang06} reported that one of the such tests would be search
of X-ray counterpart from RRATs. Particularly, presence of a
hot-spot thermal component in the possible X-ray spectra will mean
that the emission direction reversal and a preferred viewing
geometry are likely the agents to make a RRAT. While absence of
X-ray radiation indicates that considering pulsars are
"not-quite-dead pulsars before disappearing in the graveyard."

We suggest to investigate the power spectrum of the single-pulse
observational data thoroughly. In our opinion, the power spectrum
should be composition of few slightly broadened areas (ranges with
peaks) instead of thin lines. Finding this kind of signature in
the power spectra of RRATs will help to asses wether the origin of
the transient nature of these objects is indeed plasmic, as
suggested by our model.

\begin{table}
\caption{The values of pulsar parameters}             
\label{table:1}      
\centering                          
\begin{tabular}{c c c c c c}        
\hline\hline                 
PSR & $P$ & $\Delta \beta$ & $\beta _{0}$ & $\delta$ & $\vartheta$  \\    
\hline                        
   J1752+2359 & 0.41 & 0.2 & 1.2 & 1 & 0.03 \\      
   J1819-1458 & 4.26 & 0.35 & 1.35 & 1 & 0.03 \\
\hline                                   
\end{tabular}
\end{table}

\section*{Acknowledgments}

This work was partially supported by by Georgian NSF Grant
ST06/4-096, Russian Foundation for the Basic Research (project
06-02-16888) and NSF (project 00-98685).


\begin{thebibliography}{}

\bibitem [Arons (1981)]{arons81}Arons J., 1981a, Plasma Astrophys., 273
\bibitem [Gogoberidze et al.(2005)]{gog05}Gogoberidze, G., Machabeli, G.Z., Melrose, D.B., \& Luo, Q. 2005,
MNRAS, 360, 669
\bibitem [Kazbegi et al.(1991a)]{kaz91a}Kazbegi A. Z., Machabeli G. Z., Melikidze G. I., Smirnova T.
V., 1991a, Afz, 34, 433
\bibitem [Kazbegi et al.(1991b)]{kaz91b}Kazbegi A. Z., Machabeli G. Z., Melikidze G. I., 1991b, MNRAS,
253, 377
\bibitem [Kazbegi et al.(1996)]{kaz96}Kazbegi A. Z., Machabeli G. Z., Melikidze G. I., Shukre C.,
1996, A\&A, 309, 515
\bibitem [Lewandowski et al. (2004)]{Lew04}Lewandowski W., Wolszczan A., Feiler G., Konacki M., \& Soltysinski T., 2004, ApJ, 600, 905
\bibitem [Lomiashvili et al. (2006)]{lom06}Lomiashvili D., Machabeli G., \& Malov I., 2006,
ApJ, 637, 1010
\bibitem [Lyutikov et al. (1999)]{lyut99}Lyutikov M, Blandford R., Machabeli G., 1999, MNRAS, 305,
338L
\bibitem [McLaughlin et al. (2005)]{Mcl05}McLaughlin M. A., Lyne A. G., Lorimer D. R., Kramer M.,
Faulkner A. J., Manchester R. N., Cordes J. M., Camilo F.,
Possenti A., Stairs I. H., Hobbs G., D'Amico N., Burgay M., \&
O'Brien J. T., 2005, (astro-ph/0511587)
\bibitem [Sturrock (1971)]{sturrock71}Sturrock P. A., 1971, ApJ, 164, 529
\bibitem [Zhang et al. (2006)]{Zhang06}Zhang B., Gil J., \& Dyks J., 2006,
(astro-ph/0601063)
\end{thebibliography}
\end{document}